\UseRawInputEncoding
\documentclass[runinaddress,nobibnotes,twocolumn,aps,prl, reprint, floatfix,nofootinbib]{revtex4-2}

\usepackage{graphicx}
\usepackage{dcolumn}
\usepackage{bm}
\usepackage{braket}
\usepackage{xcolor}

\usepackage[normalem]{ulem}
\usepackage{booktabs} 
\usepackage{svg}
\usepackage{amsmath}
\usepackage{multirow}
\usepackage{gensymb}
\usepackage[english]{babel}
\usepackage{afterpage}

\usepackage{framed}
\colorlet{shadecolor}{yellow}
\usepackage{soul}
\usepackage{float}
\usepackage{lineno}

\begin{document}


\title{Experimental signatures of a $\hat{Z}\hat{X}$ beam-splitter interaction between a Kerr-cat and transmon qubit}


%
\author{Josiah Cochran$^{1}$}
\email{josiah.cochran@austin.utexas.edu}
\author{Haley M. Cole$^{1}$}
\email{hmcole@utexas.edu}
\author{Hebah Goderya}
\author{Zhuoqun Hao}
\author{Yao-Chun Chang}
\author{Theo Shaw}
\author{Aikaterini Kargioti}
\author{Shyam Shankar}
\email{shyam.shankar@utexas.edu}
\affiliation{Chandra Department of Electrical and Computer Engineering, University of Texas at Austin, 2501 Speedway, Austin, TX 78712,
USA
}
\footnotetext[1]{These authors contributed equally to this work.}


\date{\today}

\begin{abstract}
Quantum error correction (QEC) requires ancilla qubits to extract error syndromes from data qubits which store quantum information. However, ancilla errors can propagate back to the data qubits, introducing additional errors and limiting fault-tolerance. In superconducting quantum circuits, Kerr-cat qubits (KCQs), which exhibit strongly biased noise, have been proposed as ancillas to suppress this back-action and enhance QEC performance. Here, we experimentally demonstrate a beamsplitter interaction between a KCQ and a transmon, realizing an effective $\hat{Z}_{cat}\hat{X}_{q}$ coupling that can be employed for parity measurements in QEC protocols. We characterize the interaction across a range of cat sizes and drive amplitudes, confirming the expected scaling of the interaction rate. These results establish a step towards hybrid architectures that combine transmons as data qubits with noise-biased bosonic ancillas, enabling hardware-efficient syndrome extraction and advancing the development of fault-tolerant quantum processors.
\end{abstract}

\keywords{quantum,kerr,cat,qubit,transmon,error correcton, surface code}

\maketitle

\section{\label{intro}Introduction}

Quantum computing promises to solve problems that are classically intractable, but current hardware is limited by decoherence, gate errors, and measurement infidelity~\cite{krantz_quantum_2019}. Achieving fault-tolerant quantum computation requires quantum error correction (QEC), which encodes logical qubits across multiple physical qubits to detect and correct errors without disturbing the encoded information ~\cite{terhal_quantum_2015}. In the circuit quantum electrodynamics (cQED) platform~\cite{blais_cavity_2004, blais_circuit_2021}, many QEC demonstrations use large registers of discrete-variable qubits such as the transmon~\cite{zhao_realization_2022,google_quantum_ai_and_collaborators_quantum_2025, krinner_realizing_2022}, incurring significant hardware overhead. Bosonic qubits, which encode quantum information in the infinite-dimensional Hilbert space of a harmonic oscillator, offer a compelling hardware-efficient alternative~\cite{terhal_towards_2020, joshi_quantum_2021, cai_bosonic_2021}. Examples include cat codes, GKP codes, dual-rail codes, etc. which have demonstrated both passive and active quantum error detection and correction~\cite{ofek_demonstrating_2016, gertler_protecting_2021, reglade_quantum_2024, putterman_hardware-efficient_2025, sivak_real-time_2023, brock_quantum_2025, koottandavida2024erasure,chou_demonstrating_2023}.

The Kerr-cat qubit (KCQ) is a particular example of a bosonic cat qubit that encodes quantum information in superpositions of coherent states stabilized by Kerr nonlinearity and a two-photon drive~\cite{grimm_stabilization_2020}. Unlike the related dissipative cat qubit ~\cite{lescanne_exponential_2020, berdou_one_2023}, the KCQ is stabilized by Hamiltonian dynamics which enables fast gates while preserving the noise bias intrinsic to cat states. Recent experimental work on the KCQ has demonstrated bit-flip lifetimes of $\sim$ 500 $\mu$s, noise bias $\sim 100$ and gate fidelities above $90$\% ~\cite{frattini_observation_2024, hajr_high-coherence_2024}. The noise bias can be exploited in tailored QEC codes to improve fault tolerance~\cite{darmawan_practical_2021, xu_tailored_2023}.

The biased noise property of the KCQ has also been theorized to make it an excellent ancilla qubit for QEC when paired with conventional transmons as data qubits~\cite{puri_stabilized_2019}. Ancilla qubits are critical in QEC protocols for extracting error syndromes from data qubits ~\cite{terhal_quantum_2015}, but errors on the ancilla can propagate back to the data qubits, degrading error-correction performance ~\cite{wang_threshold_2009}.
Leveraging a bosonic ancilla with strong noise bias, such as the KCQ, offers a compelling strategy to suppress back-action and enhance fault tolerance~\cite{puri_stabilized_2019}. Given that transmons are the most widely adopted qubit in superconducting quantum processors, demonstrating a high-fidelity two-qubit interaction between a transmon and a KCQ is a crucial step toward practical integration of KCQs as ancilla. 

While prior work has shown similar couplings between a dissipative cat qubit and a transmon~\cite{putterman_hardware-efficient_2025}, that scheme required the cat stabilization drive to be turned off, temporarily removing the noise bias protection.  In contrast, our interaction is realized while the stabilization drive remains on, in theory preserving the noise bias. Separately, a three-wave mixing conditional displacement gate between a Kerr-cat and a cavity has been demonstrated ~\cite{ding_quantum_2024} and other implementations of gates between two Kerr-cats are in progress~\cite{max2024APS,ke2026APS,grimm12026APS,grimm22026APS}. However, a biased-noise-compatible interaction between a KCQ and a transmon has not been demonstrated.

In this work, we demonstrate a beamsplitter interaction between a Kerr-cat qubit and a transmon that approximates a $\hat{Z}_{cat}\hat{X}_{q}$ coupling. Such a coupling may be extended to measure multi-qubit parity observables for QEC, such as in the surface code~\cite{puri_stabilized_2019}. The interaction, driven by a single tone at the frequency difference between the two modes, leverages the unique Hamiltonian stabilization dynamics of the KCQ to enable the desired $\hat{Z}_{cat}\hat{X}_{q}$ coupling. Our results establish a key building block for using the KCQ as a bosonic ancilla for syndrome extraction in fault-tolerant QEC protocols with transmon qubits.

\section{\label{system} System description and theory of the beam-splitter interaction} 
\begin{figure}[htbp] 
\centering
\includegraphics[width=\columnwidth]{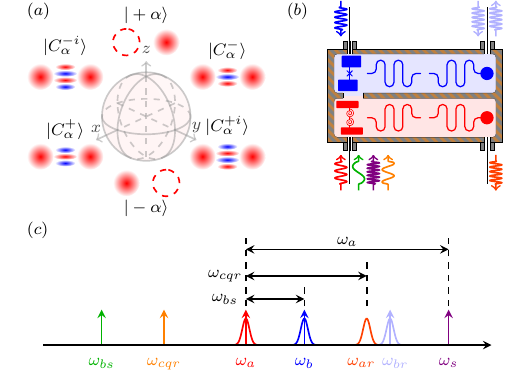}%
\caption{\label{fig1}
  (a) The Bloch sphere of the Kerr-cat with \\ even/odd parity cat states on the $x$ axis, coherent states on the $z$ axis and parity-less cat states on the $y$ axis. (b) Cartoon of the transmon and Kerr-cat with their respective drives color coded according to panel (c). Frequency spectrum of the modes and drives.
 }
\end{figure}
The architecture we employ to investigate this interaction is composed of four relevant modes, as shown in Fig.~\ref{fig1}b,c. It consists of a mode associated with the KCQ ($\omega_a$), a transmon mode ($\omega_b$) and readout resonator modes associated with the KCQ ($\omega_{ar}$) and transmon ($\omega_{br}$). 

The KCQ is realized in a driven capacitively-shunted Superconducting Nonlinear Asymmetric Inductive eLement (SNAIL) circuit~\cite{frattini_3-wave_2017}. We refer to the circuit without drive as a SNAILmon in analogy with the transmon. Ignoring the readout resonator and treating the transmon as a two-level system, the relevant terms of the system Hamiltonian can be written as 
\begin{equation}
\begin{split}
    \hat{H}_\text{sys}/\hbar=\omega_a\hat{a}^\dagger\hat{a}-K_a\hat{a}^{\dagger2}\hat{a}^2
    +\frac{\omega_b}{2}\hat{Z}_{q}
    +\frac{\chi_{ab}}{2}\hat{a}^\dagger \hat{a}\hat{Z}_{q} \\
    +g_3\left(\hat{a}+\hat{a}^\dagger\right)^3+\tilde{g_3}\left(\hat{a}^{\dagger2}\hat{\sigma}_- + \hat{a}^{2}\hat{\sigma}_+\right),
\end{split}
\label{Hsys}
\end{equation}
where $\hat{a}$ is the annihilation operators for the SNAILmon, $\hat{Z}_q$, $\hat{\sigma}_+$, $\hat{\sigma}_-$ are Pauli operators for the transmon, $2K_a$ is the SNAILmon anharmonicity, $\chi_{ab}$ is the dispersive coupling between SNAILmon and transmon and $g_3$, $\tilde{g_3}$ are terms arising from the third-order nonlinearity of the SNAIL potential.

Similar to previous experiments~\cite{frattini_squeezed_2024,grimm_stabilization_2020,ding_quantum_2024}, the KCQ is realized by applying a squeezing drive to the SNAILmon at $\omega_s=2\omega_a$, resulting in an effective KCQ Hamiltonian in the frame rotating at $\omega_a$, $\hat{H}_\text{KCQ}/\hbar=-K_a\hat{a}^{\dagger 2}\hat{a}^2 + \epsilon_2\hat{a}^{\dagger2}+\epsilon_2^*\hat{a}^2$, where $\epsilon_2$ is the proportional to $g_3$ and the amplitude of the squeezing drive~\cite{puri_stabilized_2019,grimm_stabilization_2020}.  The ground states of the KCQ Hamiltonian form a degenerate manifold spanned by even- and odd-parity cat states $\ket{\mathcal{C_\alpha^\pm}}=N_\alpha^{\pm}(\ket{\alpha}\pm \ket{-\alpha})$ with $N_\alpha$ being a normalization factor. $\epsilon_2$ and $K_a$ set the size of the cat, $\alpha^2=\epsilon_2/K_a$ and we assume $\alpha$ is a real number. This degenerate ground state manifold gives rise to the Bloch sphere, shown in Fig.~\ref{fig1}a, with cat states $\ket{\mathcal{C_\alpha^\pm}}$ chosen to lie along the $x$-axis and the coherent states $\ket{\pm\alpha}$ chosen to lie along the $z$-axis. 

The noise-biased property of the KCQ is a consequence of this encoding: single-photon loss of the SNAILmon at rate $1/T_1$ changes the parity of the oscillator which results in phase-flips $\ket{\mathcal{C_\alpha^+}}\leftrightarrow\ket{\mathcal{C_\alpha^-}}$. Thus, phase flips of the KCQ occur at the rate $1/T_c = 2\braket{\bar n}/T_1$, with $\braket{\bar n} = |\alpha|^2\left(1+e^{-4|\alpha|^2}\right)/\left(1-e^{-4|\alpha|^2}\right)$, which in the limit of large $|\alpha|$ is amplified by a factor linear in $|\alpha|^2$ compared to the photon loss rate of the SNAILmon~\cite{frattini_squeezed_2024}. Meanwhile, dephasing of the SNAILmon at rate $1/T_\phi$, cause bit-flip errors in the cat encoding. The separation of the states $\ket{\pm\alpha}$ in phase space, results in an exponentially smaller bit flip rate $1/T_\alpha = \left(|\alpha|^2/\sinh{2|\alpha|^2}\right)/T_\phi$~\cite{mirrahimi_dynamically_2014}. In practice, for moderate $|\alpha|\gtrsim 2$, the bit flip rate is set by other mechanisms such as a finite excitation rate of the SNAILmon with the exact value depending on the experiment setup~\cite{frattini_squeezed_2024}.


To create the $\hat{Z}_{cat}\hat{X}_{q}$ interaction, we apply a beam-splitter drive at the difference frequency between the KCQ and transmon ($\omega_{bs}=\omega_b-\omega_a$) and with phase $\phi$. The static effective Hamiltonian of the system in the rotating frame of the SNAILmon ($\omega_a$) and transmon ($\omega_b$) becomes 
\begin{equation}
    \hat{H}/\hbar=\hat{H}_\text{KCQ}/\hbar-\frac{\chi_{ab}}{2}\hat{a}^\dagger \hat{a}\hat{Z}_{q}+\tilde{g}_3\xi(\hat{a}^\dagger\hat{\sigma}_-e^{i\phi}+\hat{a}\hat{\sigma}_+e^{-i\phi}),
\label{rwa}
\end{equation}
where $\xi$ is the beam-splitter drive amplitude.
\begin{figure*}[htbp]
    \centering
    \includegraphics[width=0.95\textwidth]{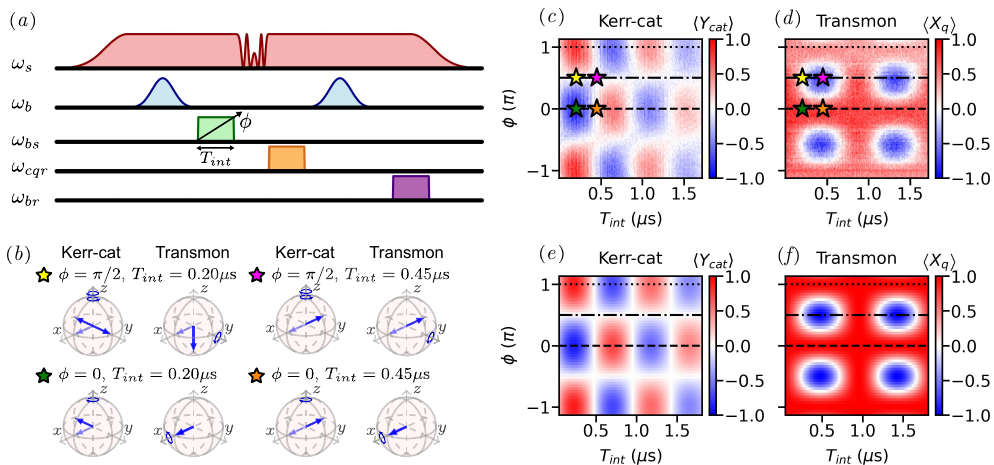}%
    \caption{\label{fig2} (a) Pulse sequence for measurement of beam-splitter interaction. The modulated region of $\omega_s$ is to perform an $\hat{X}_{cat}$ rotation on the Kerr-cat as describe in Appendix B. (b) Bloch spheres that represent the final state (light blue arrows represent the initial state) of the KCQ and transmon for different interaction time and phase (colored stars). (c, d) Experimental and (e, f) simulated signatures of the beam-splitter interaction for varying interaction time and phase for $\alpha =1.3$ and $\xi=2.6$. Colored stars in panel (c, d) correspond to the Bloch spheres in panel (b).}
    \label{fig:wide_caption} 
\end{figure*} 
Following Ref.~\cite{ding_quantum_2024}, we assume the KCQ remains in the ground-state manifold and project into the cat-qubit subspace with the projector $\mathcal{P}_C = \ket{C_\alpha^+}\bra{C_\alpha^+}+\ket{C_\alpha^-}\bra{C_\alpha^-}$. Moreover, through simulations described in Appendix C~\ref{simulations}, we find that the effect of the dispersive shift $\chi_{ab}$ is negligible. Thus, the effective interaction Hamiltonian for $\alpha\gg1$, derived in detail in Appendix C~\ref{simulations}, is given as 
\begin{equation}\label{hintg}
    \hat{H}_\text{int}/\hbar \approx \tilde{g}_3\xi\alpha \hat{Z}_{cat}(\cos{(\phi)}\hat{X}_{q}+\sin{(\phi)}\hat{Y}_{q}),
\end{equation}
where $\hat{Z}_{cat}$ is an operator on the Kerr-cat Hilbert space shown in Fig.~\ref{fig1}a, corresponding to $ \ket{C_\alpha^+}\bra{C_\alpha^-}+\ket{C_\alpha^-}\bra{C_\alpha^+}$~\cite{ding_quantum_2024}. In the rest of the text, for all Hamiltonians written with the KCQ projected into the cat-qubit subspace, we write tensor products with the KCQ operators first and transmon operators second. Eqn.~\ref{hintg} shows that the beam-splitter drive realizes the desired $\hat{Z}_{cat}\hat{X}_{q}$ interaction between KCQ and transmon, with an effective interaction rate $\Omega = \tilde{g}_3\xi\alpha$. Moreover, this interaction preserves the noise bias of the KCQ, as it does not contain terms that would introduce additional bit flips of the KCQ.

\section{\label{circuit}Circuit Realization and Characterization}

A cartoon of the experimental setup is shown in Fig.~\ref{fig1}b, showing the SNAILmon and transmon with their respective readout resonators. The sample package is similar to that of Ref.~\cite{venkatraman_driven_2024}, consisting of two halves machined out of 6061 aluminum and OFHC copper, with the package modes designed to be $>1$~GHz above $\omega_s$. The package has two 3D cavities which house the transmon and SNAILmon, respectively, with a cut-out between them for capacitive coupling of the qubits, and thus finite $\chi_{ab}$ and $\tilde{g_3}$. Each qubit is fabricated through the Dolan bridge fabrication process on separate sapphire chips. Each chip also has a lithographically defined on-chip readout resonator and Purcell filter. The transmon is designed with a single junction, whereas the SNAILmon is designed with two SNAILs in series to achieve the desired values of $\omega_a$ and $K_a$.  Details of the device design and fabrication can be found in Appendix D. Magnetic flux is applied to the SNAIL loops via a solenoid magnet, set to an operating flux $\Phi_\text{ext} = 0.33 \Phi_0$, to achieve appreciable $g_3$ and $\tilde{g}_3$.  A selected list of Hamiltonian parameters and coherence properties of the qubits at the operating point are given in Tab.~\ref{table1} and a complete list is given in Tab.~\ref{table2}. In particular, for $\alpha = 1.3$, the KCQ coherent state lifetime is $T_\alpha = 25$~$\mu$s and cat state lifetime is $T_c = 2$~$\mu$s, corresponding to a noise bias $T_\alpha/T_c = 12.5$.
\begin{table}[htbp]
\centering
\begin{tabular}{l c@{\hspace{1cm}}c}
\toprule

\textbf{SNAILmon} & & \textbf{Value} \\
\midrule
\textbf{Fock basis} & &  \\
Mode frequency & $\omega_{a}/2\pi$ & 5.2 GHz \\
Relaxation time      & $T_1$ & 40 $\mu$s \\
Ramsey decay time         & $T_{2R}$ & 5 $\mu$s \\
Operating flux       & $\Phi_\text{ext}$ &0.33 $\Phi_0$ \\
Anharmonicity        & $K_a/2\pi$ & 0.7 MHz \\
\textbf{Kerr-cat basis ($\alpha = 1.3$)}     & &  \\
Coherent state lifetime       & $T_\alpha$ & 25 $\mu$s \\
Cat state lifetime      & $T_{c}$ & 2 $\mu$s \\
\midrule
\textbf{Transmon} & & \textbf{Value} \\
\midrule
Mode frequency       & $\omega_{b}/2\pi$ & 6.7 GHz \\
Relaxation time      & $T_1$ & 33 $\mu$s \\
Ramsey decay time         & $T_{2R}$ & 47 $\mu$s \\
Hahn echo decay time      & $T_{2E}$ & 52 $\mu$s \\
\end{tabular}
\caption{\textbf{System parameters.} Selected Hamiltonian parameters and coherence times for the transmon, SNAILmon in the Fock basis, and the pumped KCQ in the cat-basis for $\alpha=1.3$.}
\label{table1}
\end{table} 

Stabilization, single-qubit gates and readout of the KCQ are performed with previously demonstrated methods~\cite{grimm_stabilization_2020,frattini_squeezed_2024,venkatraman_driven_2024,hajr_high-coherence_2024}. Control and readout of the qubits was performed with a Xilinx RFSoC with QICK firmware~\cite{ding_experimental_2024,stefanazzi_qick_2022}, which allowed direct digital synthesis of the necessary drives with stable phase relationships, obviating the need for complex mixer setups used in previous KCQ demonstrations~\cite{grimm_stabilization_2020,frattini_squeezed_2024,venkatraman_driven_2024,hajr_high-coherence_2024,ding_experimental_2024}. Details of the experimental setup can be found in Appendix D. The complete procedure for tuning up the $\hat{Z}_{cat}\hat{X}_{q}$ experiment is given in Appendix B.


\section{\label{interaction}Signatures of $\hat{Z}_{cat}\hat{X}_{q}$ interaction} 

We demonstrate the beam-splitter interaction between the KCQ and transmon using the pulse sequence shown in Fig.~\ref{fig2}a. 
The KCQ is initialized in  $\ket{\mathcal{C_\alpha^+}}$ state by turning on the squeezing drive, $\omega_s$ and the transmon is initialized in $\ket{X}$ with a $Y_{q,\pi/2}$ pulse at $\omega_b$.  Next the beam-splitter interaction is applied with a drive at $\omega_\text{bs}$ for varying time and phase. Finally, the KCQ is measured along $Y$ with a $X_{q,\pi/2}$ gate~\cite{hajr_high-coherence_2024} and cat-quadrature readout~\cite{grimm_stabilization_2020}, while the transmon is measured along $X$ with a $Y_{q,\pi/2}$ pulse and dispersive readout.


Experimental measurements and corresponding simulations as a function of the beam-splitter phase ($\phi$) and interaction time ($T_{int}$) are presented in Fig.~\ref{fig2}b-f, with the system starting in $\ket{\mathcal{C_\alpha^+}}\ket{X}$ . From Eq.~\ref{hintg}, we see that for $\phi=0,\pi$, $\hat{H}_{int}\sim\pm\hat{Z}_{cat}\hat{X}_{q}$. Thus the transmon state $\ket{X}$ is left unchanged, while the KCQ rotates around $\pm\hat{Z}_{cat}$.  The dashed ($\phi=0$) and dotted ($\phi=\pi$) lines in Fig.~\ref{fig2}c,d,e,f show the corresponding results, while Fig.~\ref{fig2}b shows the final state of the qubits on the Bloch sphere after a quarter (green star) and half (orange star) rotation for $\phi = 0$. We see that the transmon remains unchanged as a function of interaction time while the KCQ rotates clockwise for $\phi=0$ and counter clockwise for $\phi=\pi$. Next, for $\phi = \pi/2$, Eq.~\ref{hintg} implies that $\hat{H}_{int}\sim-\hat{Z}_{cat}\hat{Y}_{q}$. Thus, the transmon rotates counter-clockwise in the $x-z$ plane, as seen along the dotted-dashed line in Fig.~\ref{fig2}d,f. At the same time, the KCQ rotates around $\pm\hat{Z}_{cat}$ with equal weight, implying that $\braket{Y_{cat}}$ always evaluates to $0$ as shown by the dotted-dashed line in Fig.~\ref{fig2}c,e.

\begin{figure}[htbp]
\centering
\includegraphics[width =\columnwidth]{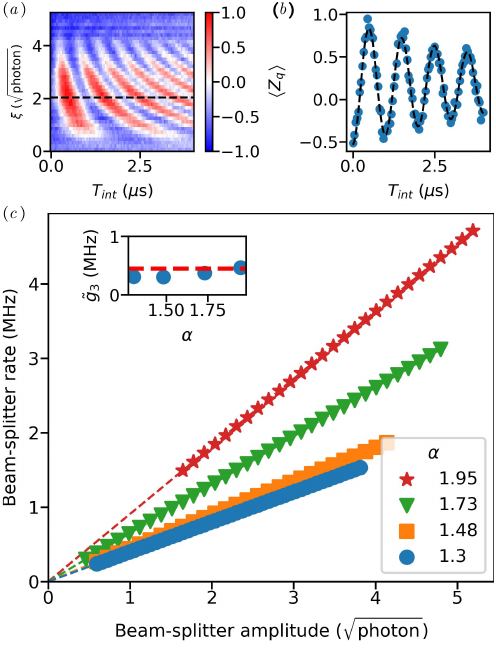}%
\caption{\label{fig3}
(a) Oscillations of the transmon qubit $\braket{Z_q}$ observable versus interaction time and drive amplitude $\xi$ when starting from $\ket{\psi} = \ket{\mathcal{C}_\alpha^+}\ket{+Z_q} = \ket{+X_{cat}}\ket{+Z_q}$, with $\alpha = 1.73$. (b) Data and fit at drive amplitude $\xi = 2.04$ (dashed line in (a)). (c) Beam-splitter rate versus drive amplitude $\xi$ for four cat sizes $\alpha$. Colored dashed lines represent the fitted value of $\tilde{g}_3\alpha\xi$ versus the drive amplitude $\xi$ (Inset) Third-order nonlinearity $\tilde{g}_3$ extracted from experiment (circles) and designed value (dashed red line).
}
\end{figure}

Next, we studied the interaction speed as a function of beam-splitter amplitude, $\xi$, and cat size, $\alpha$.  The readout contrast and coherence times for the KCQ  depends on $\alpha$, which makes fits to oscillations of the KCQ $\braket{Y_{cat}}$ observable to be less reliable as a function of $\alpha$. Instead we modified the pulse sequence to observe oscillations of the transmon state during the beam-splitter interaction. The KCQ is again initialized in $\ket{\mathcal{C_\alpha^+}}$, while the transmon is initialized in $\ket{+Z}$, such that the beam-splitter interaction results in oscillations of the transmon $\braket{Z_q}$ observable.  Fig.~\ref{fig3}a shows the transmon $\braket{Z_q}$ measurement, which oscillates with increasing rate as the beam-splitter amplitude is increased. 
The beam-splitter drive amplitude, $\xi$, was calibrated in photon number units via a separate Stark shift experiment, described in Appendix~A. A representative line cut and fit for amplitude of $\xi = 2.04$ is shown in Fig.~\ref{fig3}b. The beam-splitter drive induces an amplitude dependent Stark shift on the transmon which causes the loss of contrast for $\xi\lesssim1$ and $\xi\gtrsim4$ seen in Fig.~\ref{fig3}a. As described in Appendix E, the oscillation rate is also not linear in $\xi$ due to the frequency detuning caused by the Stark shift. Therefore, a fit which included the frequency detuning was performed and the value of the beam-splitter interaction rate $\Omega =  \tilde{g}_3\xi\alpha$ was extracted, as outlined in Appendix E. We observe that $\Omega$ varies linearly on $\xi$ as expected. According to Eqn.~\ref{hintg}, fitting the slope of this dependence and scaling by $\alpha$ gives the third-order nonlinearity $\tilde{g}_3$. The inset of Fig~\ref{fig3}c shows the measured $\tilde{g}_3$ for various $\alpha$ and compares to the designed value of $0.45$~MHz, showing excellent agreement.


Finally, we extracted the decoherence times experienced by the transmon and KCQ under the beam-splitter interaction. The decay time of the transmon $\braket{X_q}$ oscillations (shown in Fig~\ref{fig:wide_caption}d, see line cuts in Fig.~\ref{KC}a) is approximately $10$~$\mu$s, which is consistent with a master equation simulation (described in Appendix C and Fig.~\ref{KC}a). To further experimentally characterize decoherence during the interaction, we fit line cuts from Fig.~\ref{fig3} (where the transmon is now starting in $\ket{+Z_q}$) and extracted the decay time of the transmon $\braket{Z_q}$ oscillations versus the cat size and the interaction drive amplitude. The results of this fitting are presented in Fig~\ref{KC}b. In this more comprehensive data set, we observe a decrease in the decay time with increasing beam-splitter drive amplitude, while the effect of cat size remains negligible in the range studied. Moreover, the measured decay time of the transmon under all drive amplitudes, is shorter than the intrinsic $T_1$ and $T_{2R}$ of the transmon (Table~\ref{table1}). As discussed in in Appendix D, this effect is possibly due to qubit heating by noise coming down the drive line, similar to behavior observed in Refs.~\cite{ding_quantum_2024, adinolfi2025enhancingkerrcatqubitcoherence}. Understanding the cause of this effect is a subject of future study as the ability to perform transmon-KCQ gates without affecting the transmon coherence is crucial to be able to use a KCQ as an ancilla for QEC.  

\section{\label{conclusion}Conclusions and outlook}
In this work, we have experimentally demonstrated a controllable $\hat{Z}_{cat}\hat{X}_{q}$ interaction between a Kerr-cat qubit and a transmon, establishing a key building block for integrating noise-biased bosonic ancillas into error-correction protocols that use transmons as data qubits. Our results confirm that the interaction rate scales in the expected manner with both cat size and drive amplitude across a broad parameter range. Immediate next steps include investigating the limits on qubit coherence during the interaction. Recent work~\cite{adinolfi2025enhancingkerrcatqubitcoherence, ding_quantum_2024} has highlighted the role of heating  in Kerr-cat qubits, and showed that heating can degrade coherence times in both the Kerr-cat and coupled systems. Understanding and mitigating these effects is a prerequisite for accurate gate-fidelity characterization using techniques such as randomized benchmarking and gate-set tomography~\cite{hajr_high-coherence_2024}. In the medium term, our work opens a promising path toward implementing multi-qubit parity measurements with Kerr-cat ancillas, paving the way for quantum error-correction protocols that mitigate ancilla-induced back-action and enhance overall fault tolerance.

\section{Acknowledgements}
The authors thank Dr. Chao Zhou and Prof. Michael Hatridge for providing a custom version of QICK firmware that enabled this experiment. The authors also thank Dr. Nicholas Frattini, and Dr. Benjamin Brock for helpful discussion. This work was supported by the Air Force Office of Scientific Research (Grant No. FA9550-22-1-0203), the Army Research Office (Grant No. W911NF-23-1-0251 and W911NF-23-1-0096) and the Defense Advanced Research Projects Agency (Grant No. HR00112420343). HMC was supported in part by an appointment to the Department of Defense (DOD) Research Participation Program
administered by the Oak Ridge Institute for Science and Education (ORISE) through an interagency agreement between the U.S.
Department of Energy (DOE) and the DOD. ORISE is managed by Oak Ridge Associated Universities (ORAU) under DOE contract number DE-SC0014664. Sample fabrication was performed in the University of Texas at Austin Microelectronics Research Center, a member of the National Nanotechnology Coordinated Infrastructure (NNCI), which is supported by the National Science Foundation (Grant No. ECCS-2025227). All opinions expressed in this paper are the author's and do not necessarily reflect the policies and views of DOD, DOE, or ORAU/ORISE.

\section{Data Availability}

The data that support the findings of this article are
openly available~\cite{cole_2026_18168977}.

\section{\label{starkapp}Appendix A: Beam-splitter drive amplitude calibration}

To calibrate beam-splitter drive amplitude $|\xi|$, expressed in the RFSoC control as DAC units, to units of $\sqrt{\text{photons}}$ at the plane of the cavity input port, we perform a Stark shift measurement on the unpumped SNAILmon, i.e.\ in the Fock basis. When a tone is applied to the SNAILmon, the nonlinear term $K_a\hat{a}^{\dagger2}\hat{a}^2$ in its Hamiltonian becomes $K_a\vert \xi \vert^2\hat{a}^{\dagger}\hat{a}$, resulting in a frequency shift that is drive amplitude dependent. 
\begin{figure}[htbp]
\centering
\includegraphics{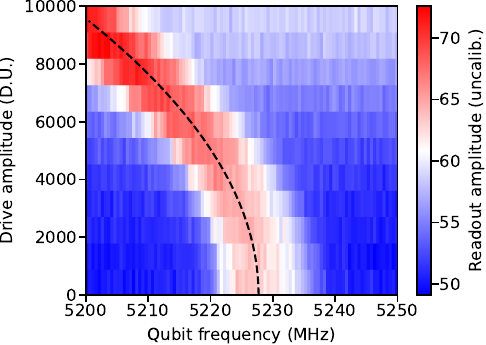}%
\caption{\label{stark}
Two tone qubit spectroscopy of the SNAILmon in the Fock-basis with a CW pulse applied at $\omega_{bs}$ with varying amplitude in DAC units (D.U.).
}
\end{figure}

Thus, by performing spectroscopy of the SNAILmon with the beamsplitter interaction drive applied, as a function of drive amplitude, we extract a conversion factor from DAC units to $\sqrt{\text{photons}}$ via a fit.
Fig.~\ref{stark} shows the spectroscopy data of the SNAILmon as a function of the beam-splitter drive amplitude. The dashed line is a fit to the equation $\omega_a-K_a(cV)^2$ where V is the drive amplitude in DAC units (D.U.) and $c$ is the fitted conversion factor in $\sqrt{\text{photon}}$/D.U. From the fit, we obtain $c = 6.57 \times 10^{-4}$ $\sqrt{\text{photon}}$/D.U.

\section{\label{C}Appendix B: Procedure for tuning the $\hat{Z}_{cat}\hat{X}_{q}$ interaction}

\begin{figure}[htbp]
\centering
\includegraphics{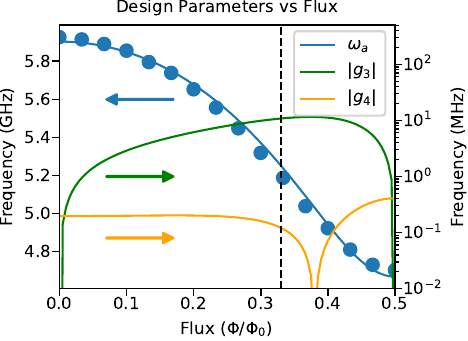}%
\caption{\label{mod}
  SNAILmon frequency versus flux (blue dots) plotted with a fit (blue line). The extracted nonlinearities $g_3$ (green), $g_4$ (orange) are calculated from the fit. Dashed line indicates the chosen operating flux of $0.33 \Phi_0$.
 }
 \end{figure}

The tuning procedure began with  continuous-wave (CW) single tone readout spectroscopy, followed by two-tone spectroscopy to determine the zero-flux frequency of the SNAILmon. Next, a two tone spectroscopy versus flux was performed, shown in Fig.~\ref{mod}. 
 \\
The flux sweep was fit to the SNAILmon model found in Ref.~\cite{nicholas_e_frattini_three-wave_2021}, to extract the charging energy $E_C$ and the linear inductive energy $E_L$. Inputs to the fit included the SNAIL asymmetry parameter of 0.1, the number of SNAILs (two) and the measured room temperature resistance values. From the fits we computed $g_3$ and $g_4$ as functions of flux, shown in Fig.~\ref{mod}. 
\begin{figure*}[htbp]
\centering
\includegraphics{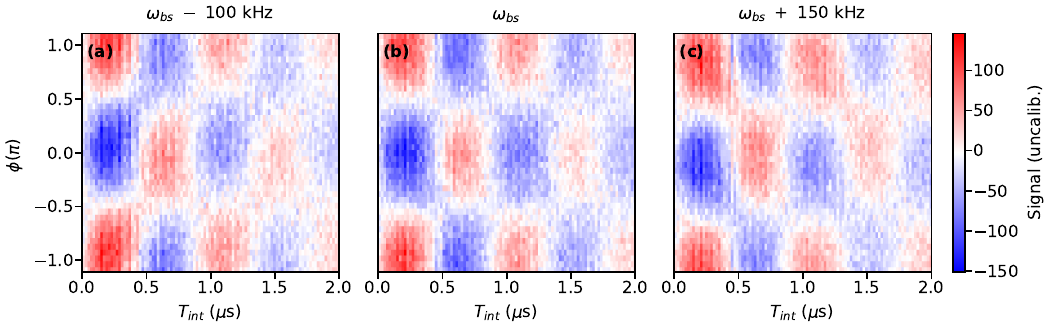}%
\caption{\label{bstune}
  Kerr-cat signal versus interaction time for three cases: (a) $\omega_{bs}$ detuned by $-100$~kHz, (b) correctly tuned, (c) detuned by $+150$~kHz.
 }
\end{figure*}
\\
\\
Since $g_3$ reaches a maximum beyond the Kerr-free point where $g_4$ changes sign, we selected an operating flux of $0.33 \Phi_0$ to maximize $g_3$ while avoiding the region of rapidly varying $g_4$. Maximizing $g_3$ is important not only to the Kerr-cat but to the interaction since the interaction arises from the three-wave mixing term. Reducing $g_3$ greatly would require a larger interaction amplitude and reduce the gate speed, introducing experimental drift.

After setting the operating point, we refined the measurement of the unpumped SNAILmon frequency and its anharmonicity using two-tone spectroscopy with reducing power. We then characterized coherence times ($T_1$, $T_2$). Next, we calibrated Kerr-cat Rabi oscillations with a Fock basis readout following Grimm et al.~\cite{grimm_stabilization_2020}. Fitting these oscillations provided the phase offset between the Rabi and squeezing drives on the RFSoC, enabling calibration of the photon number in the cat state as a function of squeezing drive amplitude.


Readout of the Kerr-cat qubit employed the cat-quadrature method~\cite{grimm_stabilization_2020,hajr_high-coherence_2024,venkatraman_driven_2024}. The cat-quadrature readout frequency $\omega_{cqr}$ was set to $\omega_a-\omega_{ar}$, where $\omega_a$ and $\omega_{ar}$ were previously determined from spectroscopy. A pulse at $\omega_{cqr}$ displaced the readout resonator proportionally to the Kerr-cat state. We optimized this pulse by varying its length and amplitude. Implementing cat-quadrature readout required an $X$-rotation, which we tuned using the phase-modulation method of Ref.~\cite{hajr_high-coherence_2024}. Once an $X(\pi/2)$ gate was calibrated, we executed the interaction pulse sequence given in Fig.~\ref{fig:wide_caption}a. 


Finally, we fine-tuned the beamsplitter frequency $\omega_{bs}$, which must equal $\omega_b - \omega_a$. This was achieved by adjusting $\omega_{bs}$ and recording Kerr-cat oscillations. Fig.~\ref{bstune} shows Kerr-cat oscillations versus interaction time for $\omega_{bs}$ at the correct frequency and with detunings of $-100$~kHz and $+150$~kHz. Negative detuning (panel a) skews the oscillations left, while positive detuning (panel c) skews them right, such that there is no phase with signal amplitude of zero. At the correct frequency (panel b), the response exhibits clear horizontal divisions at $\pm90^\circ$ phase, indicating proper tuning.

\section{\label{simulations}Appendix C: System Hamiltonian and Simulations}
In this section, we describe the system Hamiltonian following the derivations given in Refs.~\cite{nicholas_e_frattini_three-wave_2021,ding_quantum_2024}. The system Hamiltonian is written as
\begin{equation}
H_{\mathrm{total}} = H_{\mathrm{SNAIL}} + H_{\mathrm{transmon}} + H_{\mathrm{coupling}} + H_{\mathrm{drive}},
\end{equation}
where the SNAIL mode is described by
\begin{equation}
\frac{H_{\mathrm{SNAIL}}}{\hbar} = \omega_a \hat{a}^\dagger \hat{a} + \sum_{n>2} g_n (\hat{a} + \hat{a}^\dagger)^n,
\end{equation}
the transmon by
\begin{equation}
\frac{H_{\mathrm{transmon}}}{\hbar} = \omega_b \hat{b}^\dagger \hat{b} + \frac{\beta}{2} \hat{b}^{\dagger 2} \hat{b}^2,
\end{equation}
with anharmonicity 
$\beta$, and the capacitive coupling between the two modes is
\begin{equation}
\frac{H_{\mathrm{coupling}}}{\hbar} = g (\hat{a} + \hat{a}^\dagger)(\hat{b} + \hat{b}^\dagger),
\end{equation}
where $g$ is the linear coupling strength. The system is driven according to
\begin{equation}
\frac{H_{\mathrm{drive}}}{\hbar} =
\epsilon_2 e^{-i \omega_s t} \hat{a}^\dagger + \epsilon_2^* e^{i \omega_s t} \hat{a}
+ \xi e^{-i \omega_{bs} t} \hat{a}^\dagger + \xi^* e^{i \omega_{bs} t} \hat{a},
\end{equation}
where 
$\omega_s=2\omega_a$ is the squeezing drive frequency and $\omega_{bs}=\omega_a-\omega_b$ is the beam-splitter drive frequency.


To obtain an effective model,  we expand the nonlinear potential of the SNAIL mode and truncate at fourth order. Treating first the squeezing drive in $H_\textrm{drive}$, we transform to a frame rotating at $\omega_a$ and $\omega_b$ and apply a rotating wave approximation.  The Hamiltonian becomes
\begin{equation}
\begin{split}
H/\hbar = \Delta_s \hat{a}^\dagger \hat{a}
- K_a \hat{a}^{\dagger 2} \hat{a}^2 - K_b\hat{b}^{\dagger 2} \hat{b}^2
+ \left( \epsilon_2 \hat{a}^{\dagger 2} + \epsilon_2^* \hat{a}^2 \right)
\\
+ \tilde{g}_3 \left( \hat{a}^{\dagger2 } \hat{b}+ \hat{a}^2\hat{b}^\dagger \right) 
- \chi_{ab} \hat{a}^\dagger \hat{a} \hat{b}^\dagger\hat{b}.
\end{split}
\end{equation}

Here, $\epsilon_2$ denotes the effective two-photon drive amplitude generated by the squeezing tone, and $\Delta_s$ is the effective detuning in the rotating frame, including drive-induced frequency shifts.


Defining $\Delta=\omega_a-\omega_b$ and assuming the dispersive regime $g\ll\Delta$, we obtain expressions for the nonlinearities $K_a$, $K_b$, $\chi_{ab}$ and $\tilde{g}_3$ using second-order perturbation theory. The SNAILmon Kerr nonlinearity is
\begin{equation}
K_a = -6 g_4 + 30 \frac{g_3^2}{\omega_a}+\frac{\beta}{2}\left(\frac{g}{\Delta}\right)^4,
\end{equation}
where the second term arises from virtual processes mediated by the cubic nonlinearity $g_3$ and the final term originates from the transmon anharmonicity after hybridization. Similarly,
the transmon Kerr is
\begin{equation}
    K_b=\frac{\beta}{2}-6g_4\left(\frac{g}{\Delta}\right)^4+\mathcal{O}\left( \frac{g_3^2}{\omega_a}\left(\frac{g}{\Delta}\right)^4 \right),
\end{equation}
where the leading term is the intrinsic transmon anharmonicity, with a small correction arising from the SNAIL nonlinearity.
The dispersive coupling takes the form
\begin{equation}
\chi_{ab} = 2\beta\left(\frac{g}{\Delta}\right)^2 -24 g_4 \left( \frac{g}{\Delta} \right)^2 +\mathcal{O}\left( \frac{g_3^2}{\omega_a}\left(\frac{g}{\Delta}\right)^2 \right),
\end{equation}
with contributions from both systems' nonlinearities and vanish in the limit $g\rightarrow0$.
Finally, the effective third-order nonlinearity responsible for the beam-splitter interaction is given as 
\begin{equation}\label{eq:tilde_g_3}
\tilde{g}_3=3g_3\left(\frac{g}{\Delta}\right).
\end{equation}
From a separate energy-participation ratio analysis~\cite{minev2021energyparticipationquantizationjosephsoncircuits}, we calculate that $g/\Delta \sim 6\times10^{-3}$ for our device which satisfies the dispersive regime approximation. Moreover using this estimate in Eq.~\ref{eq:tilde_g_3} give $\tilde{g}_3 \sim 0.2$~MHz in reasonable agreement with the  direct estimate of $0.45$~MHz using energy-participation ratio analysis shown in Fig.~\ref{fig3}c.

Next, we treat the beam-splitter drive in $H_\textrm{drive}$. In the displaced frame of the beam-splitter tone, the third-order nonlinearity reduces into a two-photon beam-splitter interaction. Transforming to the rotating frame and retaining only the resonant terms yields
 \begin{equation}
 \begin{split}
     \hat{H}/\hbar=\hat{H}_{KCQ}/\hbar- K_b\hat{b}^{\dagger 2} \hat{b}^2-\chi_{ab}\hat{a}^\dagger\hat{a}\hat{b}^\dagger\hat{b}
     \\+\tilde{g}_3\xi(\hat{a}^\dagger\hat{b} e^{i\phi}+\hat{a}\hat{b}^\dagger e^{-i\phi})
 \end{split}
 \end{equation}
 
When we project into the cat-qubit subspace with the projector $\mathcal{P}_C = \ket{C_\alpha^+}\bra{C_\alpha^+}+\ket{C_\alpha^-}\bra{C_\alpha^-}$, the annihilation and creation operators of the KCQ take the form~\cite{nicholas_e_frattini_three-wave_2021}
\begin{equation}
\begin{split}
\hat{a}\rightarrow\alpha\left(\frac{r+r^{-1}}{2}\right)\hat{Z}_{cat}-i\alpha\left(\frac{r-r^{-1}}{2}\right)\hat{Y}_{cat} 
\\
\hat{a}^\dagger\rightarrow\alpha^*\left(\frac{r+r^{-1}}{2}\right)\hat{Z}_{cat}+i\alpha^*\left(\frac{r-r^{-1}}{2}\right)\hat{Y}_{cat}
\end{split}
\end{equation}
where $r=\frac{\sqrt{1-e^{-2|\alpha|^2}}}{\sqrt{1+e^{-2|\alpha|^2}}}$.
Assuming $\alpha$ is a real number, and plugging these expressions into the Hamiltonian and only looking at the parts of the Hamiltonian that contribute to the interaction yields
\begin{equation}
\begin{split}
    \hat{H}_{int}/\hbar= -\chi_{ab}|\alpha|^2\left(\left(B^2+C^2\right)\hat{I}_{cat}-2BC\hat{X}_{cat}\right)\hat{b}^\dagger \hat{b}
    \\
    +\tilde{g}_3\xi\alpha\left(B\hat{Z}_{cat}\left(\hat{b}e^{i\phi}+\hat{b}^\dagger e^{-i\phi}\right)+iC\hat{Y}_{cat}\left(\hat{b}e^{i\phi}-\hat{b}^\dagger e^{-i\phi}\right)\right)
    \end{split}
\end{equation}
where $B=\left(\frac{r+r^{-1}}{2}\right)$ and $C=\left(\frac{r-r^{-1}}{2}\right)$.
Treating the transmon as a two-level system, the Hamiltonian becomes
\begin{equation}
\begin{split}
\hat{H}_{int}/\hbar =
-\chi_{ab}\frac{|\alpha|^2}{2}\Biggl[
\left(\frac{r^2+r^{-2}}{2}\right)\left(\hat{I}_{cat}\hat{I}_q+\hat{I}_{cat}\hat{Z}_q\right)\\
-\left(\frac{r^2-r^{-2}}{2}\right)\left(\hat{X}_{cat}\hat{I}_q+\hat{X}_{cat}\hat{Z}_q\right)\Biggr]
\\
+\tilde{g}_3\xi\alpha\Biggl[
\left(\frac{r+r^{-1}}{2}\right)\hat{Z}_{cat}\left(\cos\phi\,\hat{X}_q+\sin\phi\,\hat{Y}_q\right)\\
-\left(\frac{r-r^{-1}}{2}\right)\hat{Y}_{cat}\left(\sin\phi\,\hat{X}_q-\cos\phi\,\hat{Y}_q\right)\Biggr]
\end{split}
\end{equation}
The large $\alpha$ limit is defined as $r\approx1$ which simplifies the above Hamiltonian to 
\begin{equation}
    \hat{H}_{int}/\hbar\approx-\chi_{ab}\frac{|\alpha|^2}{2}\hat{Z}_q + \tilde{g}_3\xi\alpha\hat{Z}_{cat}\left(\cos\phi\,\hat{X}_q+\sin\phi\,\hat{Y}_q\right).
\end{equation}
The lowest value used in the experiments is $\alpha=1.3$ which corresponds to $r=0.97$, while $\alpha=1.95$ corresponds to $r=1.0$, which allows us to make the above simplification. Furthermore, $\chi_{ab}\frac{|\alpha|^2}{2}<20$~kHz for all experiments while $\tilde{g}_3\xi\alpha \sim 1$~MHz. We also verified with numerical simulations that the $\chi_{ab}$ term has negligible effect for all values of $\alpha$ studied in the experiment.

To model the dynamics of the Kerr-cat transmon system, we solved the Lindblad master equation in Qutip~\cite{Johansson_2012} with the transmon treated as a two-level system and the Kerr-cat Hilbert space truncated to size $N=30$. The Kerr-cat basis is defined by the states on the Bloch sphere (Fig.~\ref{fig1}a as $\ket{\pm X_{cat}}=\ket{\mathcal{C_\alpha^\pm}}=\mathcal{N}_\alpha^\pm (\ket{\alpha}\pm \ket{-\alpha})$ and $\ket{\pm Z_{cat}}=\ket{\pm\alpha}$, where $\mathcal{N}_\alpha^\pm = 1/\sqrt{2(1\pm e^{-2|\alpha|^2})}$. We simulated the Hamiltonian in Eq. 2, without the $\chi_{ab}$ term, which we separately verified has negligible effect. Thus the simulated Hamiltonian is
\begin{equation}
    \hat{H}/\hbar=-K_a\hat{a}^{\dagger 2}\hat{a}^2 + \epsilon_2(\hat{a}^{\dagger2}+\hat{a}^2)+g(t)(\hat{a}\hat{\sigma}_+e^{i\phi}+\hat{a}\hat{\sigma}_-e^{-i\phi}).
\end{equation}
$\hat{a}$ is the lowering operator for the Kerr-cat mode. $g(t) = \tilde{g}_3\xi f(t)$, where $\tilde{g}_3$ is the third-order nonlinearity, $\xi$ is the interaction drive amplitude and $f(t)$ is a time-dependent sinusoidal envelope function for turning the interaction on and off. The system evolves according to the master equation 
\begin{equation}
\begin{split}
    \dot{\hat{\rho}}=-i/\hbar[\hat{H},\hat{\rho}]+\frac{1}{T_{1,a}}\mathcal{D}[\hat{a}]\hat{\rho}+\frac{1}{T_{2R,a}}\mathcal{D}[\hat{\sigma}_{z,kc}]\hat{\rho}+
    \\
    \frac{1}{T_{1,b}}\mathcal{D}[\hat{\sigma}_-]\hat{\rho}+\frac{1}{T_{2R,b}}\mathcal{D}[\hat{\sigma}_z]\hat{\rho}
\end{split}
\end{equation}
where $T_{1,a}$, $T_{2R,a}$, $T_{1,b}$, $T_{2R,b}$ are taken from experimental measurements (Table II). $\hat{Z}_{cat}$ is defined as $ \ket{C_\alpha^+}\bra{C_\alpha^-}+\ket{C_\alpha^-}\bra{C_\alpha^+}$ \cite{ding_quantum_2024}. 
\section{\label{D}Appendix D: Coherence Studies}

\begin{figure*}[htbp]
\centering
\includegraphics{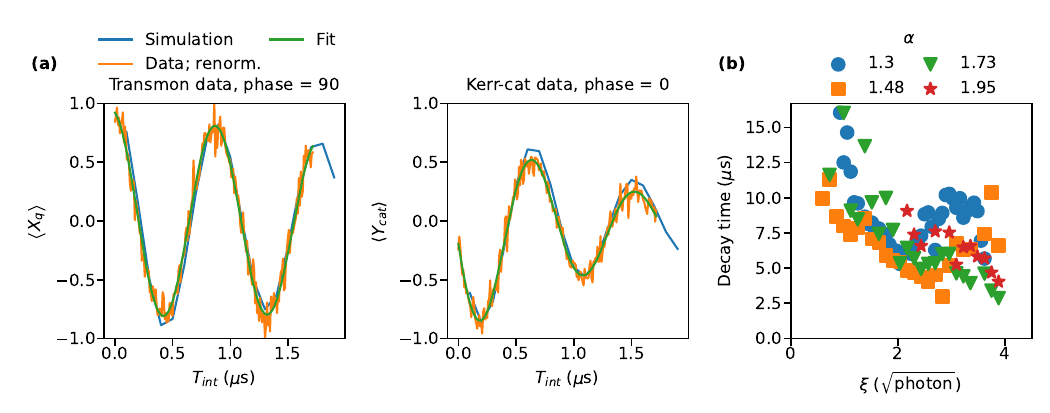}
\caption{(a) Comparison of experimental line cuts from Fig.~\ref{fig2}\textbf{c,d} at $\alpha =1.3$ and $\xi=2.6$ (orange), fits to the data (green) and simulated line cuts from Fig.~\ref{fig2}\textbf{e,f} (blue). The  transmon is initialized in $\ket{+X_q}$  and the  Kerr-cat is initialized in $\ket{+Z_{cat}}$. Transmon traces show $\braket{X_q}$ versus interaction time at a drive phase of $90\degree$, while Kerr-cat traces show $\braket{Y_{cat}}$ at phase of $0\degree$. (b) Fitted decay time of the transmon oscillations during the interaction, extracted from data in Fig.~\ref{fig3}\textbf{c} plotted as a function of interaction drive amplitude $\xi$ for several values of cat size $\alpha$. The qubits are initialized in $\ket{\psi} = \ket{\mathcal{C}_\alpha^+}\ket{+Z_q} = \ket{+X_{cat}}\ket{+Z_q}$.}
\label{KC}
\end{figure*}



Fig.~\ref{fig2}e,f shows the simulated values of the Kerr-cat $\braket{Y_{cat}}$ and transmon $\braket{X_{q}}$ observables for initial Kerr-cat,transmon state $\ket{+Z_{cat}}\ket{+X_q}$ for varying interaction time and phase. The comparison with the experimental data in Fig.~\ref{fig2}c,d indicates good agreement. Moreover, line cuts of these data and simulations are shown in Fig.~\ref{KC}a, demonstrating good agreement when dephasing and single-photon loss are included. 

However, we observe the extracted decay times decrease with increasing drive strength, while the dependence on cat size is negligible within the explored range. This reduction in decay time with $\xi$ is not captured by the master equation simulations. This effect is possibly due to qubit heating and requires further investigation. 

One possible cause for qubit heating is the wiring configuration of the experiment. In the setup shown in Fig.~\ref{wiring}, the Kerr-cat Rabi drive is combined at room temperature with other high-power three-wave mixing (3WM) drives, rather than being routed through a dedicated, independently filtered line. This choice limits the effectiveness of cold attenuation and filtering at the Kerr-cat and transmon frequency, which could result in increased thermal populations of the modes and correspondingly reduced cat state lifetime and transmon coherence time. It could also be responsible for the reduced $T_\alpha$ and $T_c$ of the device without the beam-splitter interaction applied. Similar effects from heating have been seen in Ref.~\cite{ding_quantum_2024} and studied further in Ref.~\cite{adinolfi2025enhancingkerrcatqubitcoherence}.

\section{\label{E}Appendix E: Extraction of Oscillation Rate}
\begin{figure}[H]
\centering
\includegraphics{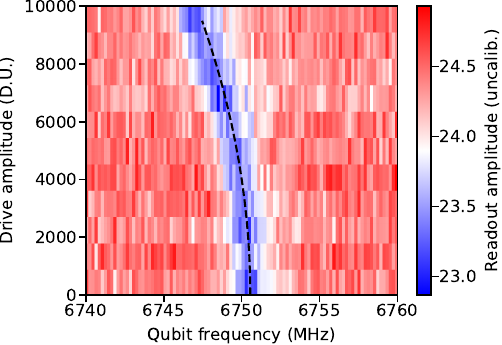}%
\caption{\label{stark_t}
Two tone qubit spectroscopy of the transmon with a CW pulse applied at $\omega_{bs}$ with varying amplitude in DAC units (D.U.).
}
\end{figure}

For the experiments, shown in Fig.~\ref{fig3}, the beam-splitter drive was set at a fixed frequency of $\omega_{bs}/2\pi = $ 1.5149~GHz, corresponding to the frequency used in the experiments shown in Fig.~\ref{fig2}c,d. However the transmon experiences a Stark shift with increasing drive amplitude, shown in Fig.~\ref{stark_t}.

\begin{figure}[htbp]
\centering
\includegraphics{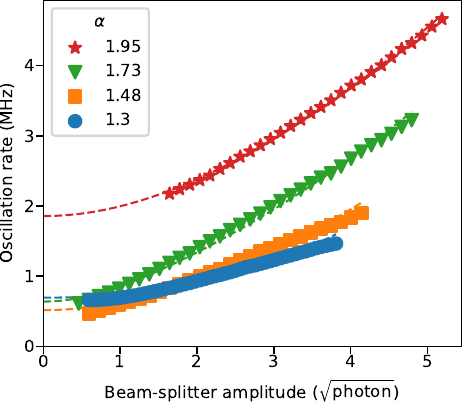}%
\caption{\label{oscillation}
Oscillation rate versus beam-splitter drive amplitude, $\xi$ for four cat sizes, $\alpha$. Colored dashed lines represent fits to model including detuning from Stark shift.
}
\end{figure}

The beam-splitter frequency $\omega_{bs}/2\pi = $ 1.5149~GHz, corresponds to an on-resonance condition for drive amplitude of 4000~D.U. At this drive amplitude, the transmon Stark shift is observed to be approximately $0.5$~MHz. At other drive amplitudes the beam-splitter drive will be detuned. Therefore to extract the beam-splitter rate we fit the oscillation rate to the form  $\sqrt{\Omega^2+\Delta^2}= \sqrt{(a\xi)^2+(b\xi^2+c)^2}$
where the beam-splitter rate is $\Omega = a\xi=\tilde{g}_3\alpha\xi$, and the detuning $\Delta = b\xi^2+c$ comprises the beam-splitter amplitude dependent Stark shift $b\xi^2$, and residual detuning $c$ to account for frequency shifts that are independent of beam-splitter amplitude. We extract the value of $b$ from the two-tone spectroscopy on the transmon shown in Fig.~\ref{stark_t}.  We then fix $b$ and fit $c$ and $a$. The oscillation rate versus $\xi$ for different cat size $\alpha$ along with their respective fits are shown in Fig.~\ref{oscillation}. The fitted values of $a=\tilde{g}_3\alpha$ are used to plot the beam-splitter rate $\Omega$ in Fig.~\ref{fig3}c.  We expect $c\sim 0.5$~MHz due to the transmon Stark-shift mentioned above, as well as a two-orders of magnitude smaller $\alpha$ dependent shift due to the Stark-shift of the Kerr-cat with the squeezing drive strength $\epsilon_2$. The fits give values of $c$ close to $0.5$~MHz for $\alpha = 1.3, 1.48, 1.73$ and an outlier value of 1.7~MHz, which is unexplained.
In Fig.~\ref{fig3}a, the oscillation contrast is seen to disappear at low and high beam-splitter amplitude. This effect also arises due to the fixed beam-splitter $\omega_{bs}$ frequency used in the experiment. From the point-of-view of the transmon, the beam-splitter interaction corresponds to a detuned Rabi oscillation. Thus the oscillation contrast goes as $\Omega^2/(\Omega^2+\Delta^2)$, which decreases as $\Delta$ increases due to the Stark-shift of the transmon.


\section{\label{F}Appendix F: Device fabrication, parameters and experimental setup details}

Both chips used in this experiment were fabricated on c-plane sapphire substrates. Wafers were cleaned with N-methyl-2-pyrrolidone (NMP), acetone, and isopropyl alcohol (IPA) followed by a 5-minute dehydration bake at 175$\degree $C. For e-beam lithography, a bilayer resist stack was applied: 650~nm MMA EL13 and 250~nm PMMA A4, baked at 175$\degree $C  (1 minute for the first layer, 30~minutes for the bilayer). To mitigate charging during lithography, a 14~nm aluminum anti-charging layer was deposited using an Angstrom e-beam evaporator. Patterns for on-chip readout resonators, Purcell filters, and Josephson junctions were written with a JEOL~8100 e-beam writer. After exposure, the aluminum layer was removed in AD-10 developer (90 s) and then the pattern was developed in 3:1 IPA:H$_2$O at 6$\degree $C for 2 minutes, followed by blow dry with nitrogen. 

The substrate was then loaded into the evaporator, pumped to below $2\times10^{-7}$ torr, and ion milled with an Ar beam for 90~s. Two aluminum layers are deposited at $\pm$30$\degree $, separated by a static oxidation step at 15 torr for 12 minutes with 85:15 Ar:O$_2$ mixture. After the second aluminum layer deposition, a surface cap oxide was grown at 3~torr for 10~minutes. Liftoff was performed in NMP at 90$\degree $C for 2~hours, followed by rinsing in fresh NMP and sonication for 1~minute. Devices were then cleaned in acetone, rinsed in IPA, dried with nitrogen, and cleaved using a SYJ-DS100-LD scribe tool.

\begin{figure*}[htbp]
\centering
\includegraphics[scale=0.85]{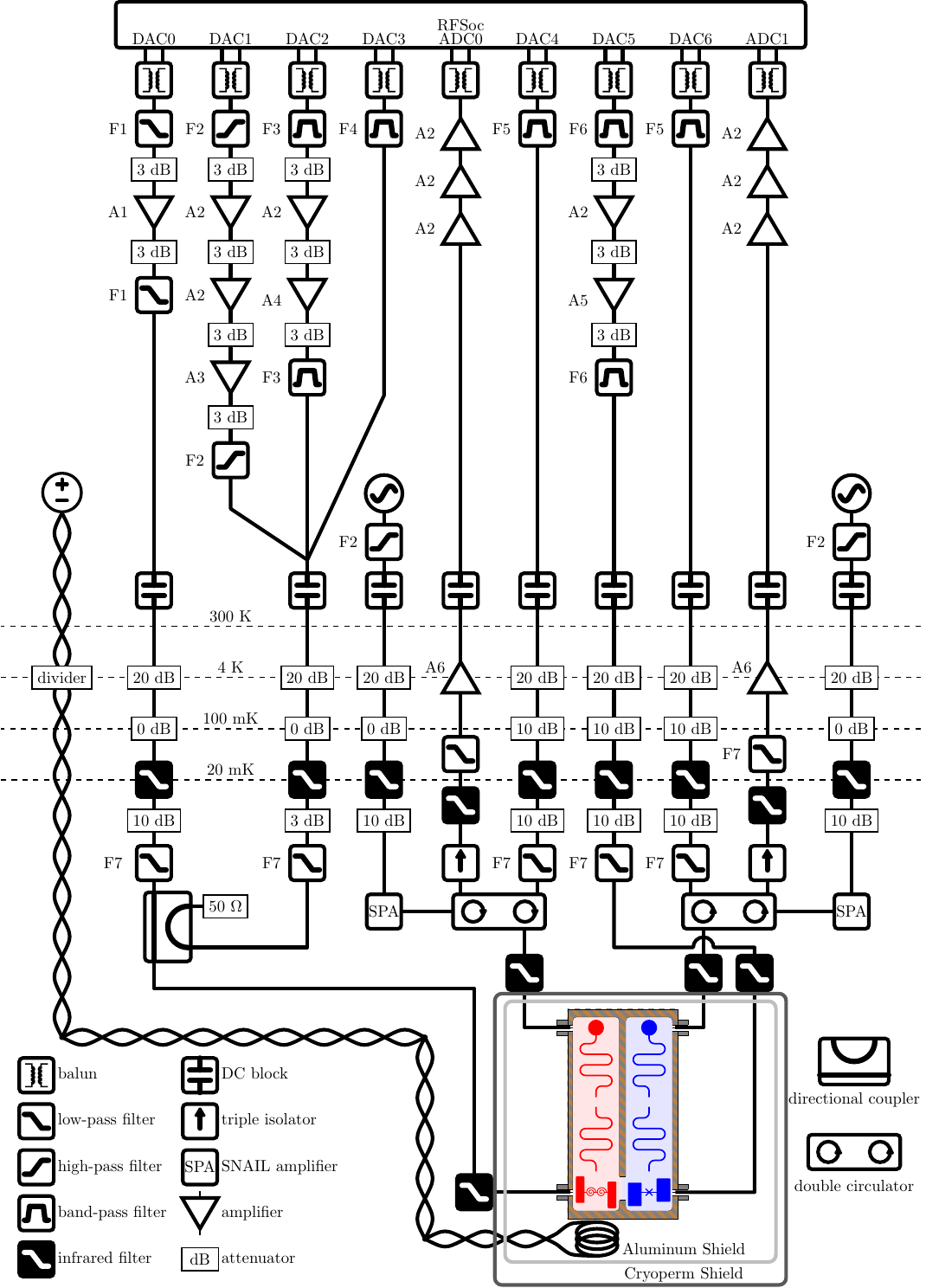}
\caption{Wiring diagram of the experimental setup, with component definitions referenced in Table~\ref{table3}. DAC0 through DAC6 provide the following drive signals in order: $\omega_{bs}$, $\omega_{s}$, $\omega_{cqr}$, $\omega_{a}$, $\omega_{ar}$, $\omega_{b}$, $\omega_{br}$. ADC0 and ADC1 digitize the readout signals from the Kerr-cat and transmon respectively.}
\label{wiring}
\end{figure*}

Measured and simulated device parameters are summarized in Table~\ref{table2}, along with the methods used to obtain them taken from Refs.~\cite{grimm_stabilization_2020,frattini_squeezed_2024,hajr_high-coherence_2024}. 
\begin{table*}[htbp]
\begin{tabular}{l c@{\hspace{1cm}}c}
\midrule
\textbf{Parameter} & \textbf{Value} & \textbf{Measurement or estimate method} \\ 
\midrule
SNAIL charging energy, $E_C/h$                           & 109 MHz          & SNAIL fit                         \\
Number of SNAILs                                         & 2         & Design                   \\
SNAIL asymmetry                            & 0.1       & Room temperature resistance measurement                         \\
SNAIL inductance, $L_J$                                  & 0.6 nH& Room temperature resistance measurement                          \\
SNAIL frequency at $\Phi/\Phi_0=0$                       & 5.93 GHz          & Two-tone spectroscopy    \\
SNAIL frequency at $\Phi/\Phi_0=0.5$                     & 4.7 GHz          & Two-tone spectroscopy    \\ 
\midrule
SNAIL operating bias, $\Phi_{\text{ext}}/\Phi_0$                      & 0.33      & Design                   \\
SNAIL operating frequency, $\omega_a/2\pi$               & 5.2 GHz          & Two-tone spectroscopy    \\
SNAIL cubic nonlinearity, $g_3/2\pi$                     & 11 MHz    & Design and SNAIL fit     \\
SNAIL self-Kerr nonlinearity, $K_a/2\pi$                   & 0.7 MHz          & Two-tone spectroscopy    \\
Fock basis relaxation time, $T_1$                        & 40 $\mu$s & Coherence measurement    \\
Fock basis Ramsey decay, $T_{2R}$                        & 5 $\mu$s  & Coherence measurement    \\
Coherent state lifetime, $T_\alpha$ for $\alpha = 1.3$                     & 25 $\mu$s & Coherence measurement                         \\
Cat state lifetime, $T_c$  for $\alpha = 1.3$                             & 2 $\mu$s  & Coherence measurement                          \\ \midrule
Transmon mode frequency, $\omega_b/2\pi$                 & 6.7 GHz   & Two-tone spectroscopy    \\
Transmon relaxation time, $T_1$                          & 33 $\mu$s & Coherence measurement    \\
Transmon Ramsey decay, $T_{2R}$                          & 47 $\mu$s & Coherence measurment     \\
Transmon Hahn echo decay, $T_{2E}$                       & 52 $\mu$s & Coherence measurement    \\ \midrule
SNAIL readout resonator frequency, $\omega_{ar}/2\pi$    & 8.3 GHz          & Single-tone spectroscopy \\
SNAIL readout resonator linewidth, $\kappa_{ar}/2\pi$    & 0.1 MHz          & Single-tone spectroscopy \\
Transmon readout resonator frequency, $\omega_{br}/2\pi$ & 8.56 GHz          & Single-tone spectroscopy \\
Transmon readout resonator linewidth, $\kappa_{br}/2\pi$ & 2.06 MHz          & Single-tone spectroscopy \\ \midrule
CQR mode frequency, $\omega_{cqr}/2\pi$                  & 3.1 GHz          & Two-tone spectroscopy    \\
Squeezing mode frequency, $\omega_{s}/2\pi$              & 10.4 GHz          & Two-tone spectroscopy    \\
Beam-splitter mode frequency, $\omega_{bs}/2\pi$         & 1.5 GHz          & Two-tone spectroscopy    \\
Kerr-cat to transmon cross-Kerr, $\chi_{ab}/2\pi$             & 10 kHz & Simulation     
\end{tabular}
\caption{Summary of device parameters discussed in the main text and appendix, along with the method used to obtain them. Design parameters were determined by Ansys HFSS and pyEPR~\cite{minev2021energyparticipationquantizationjosephsoncircuits} with corrections described in Ref.~\cite{nicholas_e_frattini_three-wave_2021}. Horizontal lines separate groups of parameters: SNAIL design values, SNAIL parameters at the operating flux, transmon frequency and coherence, readout resonator characteristics, and additional frequencies necessary for the experiment.  }
\label{table2}
\end{table*}
The experimental setup and wiring diagram are shown in Fig.~\ref{wiring}, with component part numbers listed in Table~\ref{table3}. A key feature of this setup, compared to previous Kerr-cat experiments~\cite{grimm_stabilization_2020,hajr_high-coherence_2024,venkatraman_driven_2024}, is the use of a Xilinx RFSoC in a mixer-less configuration with QICK firmware~\cite{stefanazzi_qick_2022,ding_experimental_2024}. A custom firmware version was used to enable reset of the clock phase across all DAC and ADC channels prior to pulse execution~\cite{ding_experimental_2024}. The use of this firmware with phase reset capability was crucial to ensure precise phase alignment for qubit control, readout, and beamsplitter drives during each shot of the experiment. 

\begin{table*}[htbp]
\begin{tabular}{l c@{\hspace{1cm}}l c@{\hspace{1cm}}lc}

\textbf{Filter} & \textbf{Part no.} & \textbf{Amplifier} & \textbf{Part no.} & \textbf{Balun}      & \textbf{Part no.}  \\ \midrule
F1              & ZSS2252-100W-S+   & A1                 & ZVA-1W-103+       &                     & MABA-011108        \\  
F2              & ZHSS-11G-S+       & A2                 & ZX60-123LN-S+     & \textbf{Coupler}    &                    \\ 
F3              & ZBSS-3G-S+        & A3                 & ZVE-3W-183+       &                     & QMC-CRYOCOUPLER-10 \\ 
F4              & VBFZ-5500-S+      & A4                 & ZVE-3W-83+        & \textbf{Circulator} &                    \\  
F5              & ZBSS-10G-S+       & A5                 & ZVA-183-S+        & dual junction       & LNF-CICIC4 12A S   \\
F6              & ZBSS-6G-S+        & A6                 & LNF-LNC4 16B      & triple junction     & LNF-CIISISC4 12A   \\
F7              & 6L250-12000       &                    &                   &                     &                   
\end{tabular}
\caption{Part numbers for all components used in the experiment, listed according to their labels in the wiring diagram shown in Fig.~\ref{wiring}.}
\label{table3}
\end{table*}

\clearpage

\bibliography{references}

@misc{max2024APS,
  author       = {Schaefer, Max},
  title        = {Experimental progress towards a bias-preserving CNOT gate between two Kerr-cat qubits},
  howpublished = {Presented at the APS March Meeting, Minneapolis, MN},
  month        = {March},
  year         = {2024},
  note         = {Session Q47.00010},
  url          = {https://aps.org}
}

@misc{grimm22026APS,
  author       = {Michaud, Laurent},
  title        = {A triplet of coupled Kerr-nonlinear oscillators in the quantum regime (part 2)},
  howpublished = {Presented at the APS March Meeting, Denver, CO},
  month        = {March},
  year         = {2026},
  note         = {Session J16: Kerr-Cat Qubits: Control, Gates, Readout, and Applications},
  url          = {https://summit.aps.org/smt/2026/events/MAR-S62}
}

@misc{grimm12026APS,
  author       = {Kamrul, Venus},
  title        = {A triplet of coupled Kerr-nonlinear oscillators in the quantum regime (part 1)},
  howpublished = {Presented at the APS March Meeting, Denver, CO},
  month        = {March},
  year         = {2026},
  note         = {Session J16: Kerr-Cat Qubits: Control, Gates, Readout, and Applications},
  url          = {https://summit.aps.org/smt/2026/events/MAR-S62}
}

@misc{ke2026APS,
  author       = {Wang, Ke},
  title        = {Experimental control-Z two qubit gate on 2D Kerr cats},
  howpublished = {Presented at the APS March Meeting, Denver, CO},
  month        = {March},
  year         = {2026},
  note         = {Session J16: Kerr-Cat Qubits: Control, Gates, Readout, and Applications},
  url          = {https://summit.aps.org/smt/2026/events/MAR-S62}
}

@misc{adinolfi2025enhancingkerrcatqubitcoherence,
      title={Enhancing Kerr-Cat Qubit Coherence with Controlled Dissipation}, 
      author={Francesco Adinolfi and Daniel Z. Haxell and Alessandro Bruno and Laurent Michaud and Venus Hasanuzzaman Kamrul and Preeti Pandey and Alexander Grimm},
      year={2025},
      eprint={2511.01027},
      archivePrefix={arXiv},
      primaryClass={quant-ph},
      url={https://arxiv.org/abs/2511.01027}, 
}

@dataset{cole_2026_18168977,
  author       = {Cole, Haley},
  title        = {Experimental signatures of a \$\sigma\_z\sigma\_x\$
                   beam-splitter interaction between a Kerr-cat and
                   transmon qubit
                  },
  month        = jan,
  year         = 2026,
  publisher    = {Zenodo},
  doi          = {10.5281/zenodo.18168977},
  url          = {https://doi.org/10.5281/zenodo.18168977},
}

@article{Johansson_2012,
   title={QuTiP: An open-source Python framework for the dynamics of open quantum systems},
   volume={183},
   ISSN={0010-4655},
   url={http://dx.doi.org/10.1016/j.cpc.2012.02.021},
   DOI={10.1016/j.cpc.2012.02.021},
   number={8},
   journal={Computer Physics Communications},
   publisher={Elsevier BV},
   author={Johansson, J.R. and Nation, P.D. and Nori, Franco},
   year={2012},
   month=aug, pages={1760–1772} }

@article{minev2021energyparticipationquantizationjosephsoncircuits,
  title={Energy-participation quantization of Josephson circuits},
  author={Minev, Zlatko K and Leghtas, Zaki and Mundhada, Shantanu O and Christakis, Lysander and Pop, Ioan M and Devoret, Michel H},
  journal={npj Quantum Information},
  volume={7},
  number={1},
  pages={131},
  year={2021},
  publisher={Nature Publishing Group UK London}
}

@article{wang_threshold_2009,
author = {Wang, D. S. and Fowler, A. G. and Stephens, A. M. and Hollenberg, L. C. L.},
title = {Threshold error rates for the toric and planar codes},
year = {2010},
issue_date = {May 2010},
publisher = {Rinton Press, Incorporated},
address = {Paramus, NJ},
volume = {10},
number = {5},
issn = {1533-7146},
journal = {Quantum Info. Comput.},
month = may,
pages = {456–469},
numpages = {14}
}

@article{chou_demonstrating_2023,
  title={A superconducting dual-rail cavity qubit with erasure-detected logical measurements},
  author={Chou, Kevin S and Shemma, Tali and McCarrick, Heather and Chien, Tzu-Chiao and Teoh, James D and Winkel, Patrick and Anderson, Amos and Chen, Jonathan and Curtis, Jacob C and de Graaf, Stijn J and others},
  journal={Nature Physics},
  volume={20},
  number={9},
  pages={1454--1460},
  year={2024},
  publisher={Nature Publishing Group UK London}
}

@article{koottandavida2024erasure,
  title={Erasure detection of a dual-rail qubit encoded in a double-post superconducting cavity},
  author={Koottandavida, Akshay and Tsioutsios, Ioannis and Kargioti, Aikaterini and Smith, Cassady R and Joshi, Vidul R and Dai, Wei and Teoh, James D and Curtis, Jacob C and Frunzio, Luigi and Schoelkopf, Robert J and others},
  journal={Physical Review Letters},
  volume={132},
  number={18},
  pages={180601},
  year={2024},
  publisher={APS}
}

@article{berdou_one_2023,
	title = {One {Hundred} {Second} {Bit}-{Flip} {Time} in a {Two}-{Photon} {Dissipative} {Oscillator}},
	volume = {4},
	issn = {2691-3399},
	url = {https://link.aps.org/doi/10.1103/PRXQuantum.4.020350},
	doi = {10.1103/PRXQuantum.4.020350},
	language = {en},
	number = {2},
	urldate = {2025-11-13},
	journal = {PRX Quantum},
	author = {Berdou, C. and Murani, A. and Réglade, U. and Smith, W.C. and Villiers, M. and Palomo, J. and Rosticher, M. and Denis, A. and Morfin, P. and Delbecq, M. and Kontos, T. and Pankratova, N. and Rautschke, F. and Peronnin, T. and Sellem, L.-A. and Rouchon, P. and Sarlette, A. and Mirrahimi, M. and Campagne-Ibarcq, P. and Jezouin, S. and Lescanne, R. and Leghtas, Z.},
	month = jun,
	year = {2023},
	pages = {020350},
}

@article{lescanne_exponential_2020,
	title = {Exponential suppression of bit-flips in a qubit encoded in an oscillator},
	volume = {16},
	issn = {1745-2473, 1745-2481},
	url = {https://www.nature.com/articles/s41567-020-0824-x},
	doi = {10.1038/s41567-020-0824-x},
	language = {en},
	number = {5},
	urldate = {2025-11-13},
	journal = {Nature Physics},
	author = {Lescanne, Raphaël and Villiers, Marius and Peronnin, Théau and Sarlette, Alain and Delbecq, Matthieu and Huard, Benjamin and Kontos, Takis and Mirrahimi, Mazyar and Leghtas, Zaki},
	month = may,
	year = {2020},
	pages = {509--513},
}

@article{cai_bosonic_2021,
	title = {Bosonic quantum error correction codes in superconducting quantum circuits},
	volume = {1},
	issn = {26673258},
	url = {https://linkinghub.elsevier.com/retrieve/pii/S2667325820300145},
	doi = {10.1016/j.fmre.2020.12.006},
	language = {en},
	number = {1},
	urldate = {2025-11-13},
	journal = {Fundamental Research},
	author = {Cai, Weizhou and Ma, Yuwei and Wang, Weiting and Zou, Chang-Ling and Sun, Luyan},
	month = jan,
	year = {2021},
	pages = {50--67},
}

@article{joshi_quantum_2021,
	title = {Quantum information processing with bosonic qubits in circuit {QED}},
	volume = {6},
	issn = {2058-9565},
	url = {https://iopscience.iop.org/article/10.1088/2058-9565/abe989},
	doi = {10.1088/2058-9565/abe989},
	number = {3},
	urldate = {2025-11-13},
	journal = {Quantum Science and Technology},
	author = {Joshi, Atharv and Noh, Kyungjoo and Gao, Yvonne Y},
	month = jul,
	year = {2021},
	pages = {033001},
}

@article{brock_quantum_2025,
	title = {Quantum error correction of qudits beyond break-even},
	volume = {641},
	issn = {0028-0836, 1476-4687},
	url = {https://www.nature.com/articles/s41586-025-08899-y},
	doi = {10.1038/s41586-025-08899-y},
	language = {en},
	number = {8063},
	urldate = {2025-11-13},
	journal = {Nature},
	author = {Brock, Benjamin L. and Singh, Shraddha and Eickbusch, Alec and Sivak, Volodymyr V. and Ding, Andy Z. and Frunzio, Luigi and Girvin, Steven M. and Devoret, Michel H.},
	month = may,
	year = {2025},
	pages = {612--618},
}

@article{sivak_real-time_2023,
	title = {Real-time quantum error correction beyond break-even},
	volume = {616},
	issn = {0028-0836, 1476-4687},
	url = {https://www.nature.com/articles/s41586-023-05782-6},
	doi = {10.1038/s41586-023-05782-6},
	language = {en},
	number = {7955},
	urldate = {2025-11-13},
	journal = {Nature},
	author = {Sivak, V. V. and Eickbusch, A. and Royer, B. and Singh, S. and Tsioutsios, I. and Ganjam, S. and Miano, A. and Brock, B. L. and Ding, A. Z. and Frunzio, L. and Girvin, S. M. and Schoelkopf, R. J. and Devoret, M. H.},
	month = apr,
	year = {2023},
	pages = {50--55},
}

@article{krinner_realizing_2022,
	title = {Realizing repeated quantum error correction in a distance-three surface code},
	volume = {605},
	issn = {0028-0836, 1476-4687},
	url = {https://www.nature.com/articles/s41586-022-04566-8},
	doi = {10.1038/s41586-022-04566-8},
	language = {en},
	number = {7911},
	urldate = {2025-11-13},
	journal = {Nature},
	author = {Krinner, Sebastian and Lacroix, Nathan and Remm, Ants and Di Paolo, Agustin and Genois, Elie and Leroux, Catherine and Hellings, Christoph and Lazar, Stefania and Swiadek, Francois and Herrmann, Johannes and Norris, Graham J. and Andersen, Christian Kraglund and Müller, Markus and Blais, Alexandre and Eichler, Christopher and Wallraff, Andreas},
	month = may,
	year = {2022},
	pages = {669--674},
}

@article{terhal_quantum_2015,
	title = {Quantum error correction for quantum memories},
	volume = {87},
	copyright = {http://link.aps.org/licenses/aps-default-license},
	issn = {0034-6861, 1539-0756},
	url = {https://link.aps.org/doi/10.1103/RevModPhys.87.307},
	doi = {10.1103/RevModPhys.87.307},
	language = {en},
	number = {2},
	urldate = {2025-11-13},
	journal = {Reviews of Modern Physics},
	author = {Terhal, Barbara M.},
	month = apr,
	year = {2015},
	pages = {307--346},
}

@article{blais_circuit_2021,
	title = {Circuit quantum electrodynamics},
	volume = {93},
	issn = {0034-6861, 1539-0756},
	url = {https://link.aps.org/doi/10.1103/RevModPhys.93.025005},
	doi = {10.1103/RevModPhys.93.025005},
	language = {en},
	number = {2},
	urldate = {2025-10-14},
	journal = {Reviews of Modern Physics},
	author = {Blais, Alexandre and Grimsmo, Arne L. and Girvin, S. M. and Wallraff, Andreas},
	month = may,
	year = {2021},
	pages = {025005},
}

@article{google_quantum_ai_and_collaborators_quantum_2025,
	title = {Quantum error correction below the surface code threshold},
	volume = {638},
	issn = {0028-0836, 1476-4687},
	url = {https://www.nature.com/articles/s41586-024-08449-y},
	doi = {10.1038/s41586-024-08449-y},
	language = {en},
	number = {8052},
	urldate = {2025-10-14},
	journal = {Nature},
	author = {{Google Quantum AI and Collaborators} and Acharya, Rajeev and Abanin, Dmitry A. and Aghababaie-Beni, Laleh and Aleiner, Igor and Andersen, Trond I. and Ansmann, Markus and Arute, Frank and Arya, Kunal and Asfaw, Abraham and Astrakhantsev, Nikita and Atalaya, Juan and Babbush, Ryan and Bacon, Dave and Ballard, Brian and Bardin, Joseph C. and et. al.},
	month = feb,
	year = {2025},
	pages = {920--926},
}

@article{zhao_realization_2022,
	title = {Realization of an {Error}-{Correcting} {Surface} {Code} with {Superconducting} {Qubits}},
	volume = {129},
	issn = {0031-9007, 1079-7114},
	url = {https://link.aps.org/doi/10.1103/PhysRevLett.129.030501},
	doi = {10.1103/PhysRevLett.129.030501},
	language = {en},
	number = {3},
	urldate = {2025-10-14},
	journal = {Physical Review Letters},
	author = {Zhao, Youwei and Ye, Yangsen and Huang, He-Liang and Zhang, Yiming and Wu, Dachao and Guan, Huijie and Zhu, Qingling and Wei, Zuolin and He, Tan and Cao, Sirui and Chen, Fusheng and Chung, Tung-Hsun and Deng, Hui and Fan, Daojin and Gong, Ming and Guo, Cheng and Guo, Shaojun and Han, Lianchen and Li, Na and Li, Shaowei and Li, Yuan and Liang, Futian and Lin, Jin and Qian, Haoran and Rong, Hao and Su, Hong and Sun, Lihua and Wang, Shiyu and Wu, Yulin and Xu, Yu and Ying, Chong and Yu, Jiale and Zha, Chen and Zhang, Kaili and Huo, Yong-Heng and Lu, Chao-Yang and Peng, Cheng-Zhi and Zhu, Xiaobo and Pan, Jian-Wei},
	month = jul,
	year = {2022},
	pages = {030501},
}

@article{blais_cavity_2004,
	title = {Cavity quantum electrodynamics for superconducting electrical circuits: {An} architecture for quantum computation},
	volume = {69},
	copyright = {http://link.aps.org/licenses/aps-default-license},
	issn = {1050-2947, 1094-1622},
	shorttitle = {Cavity quantum electrodynamics for superconducting electrical circuits},
	url = {https://link.aps.org/doi/10.1103/PhysRevA.69.062320},
	doi = {10.1103/PhysRevA.69.062320},
	language = {en},
	number = {6},
	urldate = {2025-10-14},
	journal = {Physical Review A},
	author = {Blais, Alexandre and Huang, Ren-Shou and Wallraff, Andreas and Girvin, S. M. and Schoelkopf, R. J.},
	month = jun,
	year = {2004},
	pages = {062320},
}

@article{ding_experimental_2024,
	title = {Experimental advances with the {QICK} ({Quantum} {Instrumentation} {Control} {Kit}) for superconducting quantum hardware},
	volume = {6},
	issn = {2643-1564},
	url = {https://link.aps.org/doi/10.1103/PhysRevResearch.6.013305},
	doi = {10.1103/PhysRevResearch.6.013305},
	language = {en},
	number = {1},
	urldate = {2025-09-23},
	journal = {Physical Review Research},
	author = {Ding, Chunyang and Di Federico, Martin and Hatridge, Michael and Houck, Andrew and Leger, Sebastien and Martinez, Jeronimo and Miao, Connie and I, David Schuster and Stefanazzi, Leandro and Stoughton, Chris and Sussman, Sara and Treptow, Ken and Uemura, Sho and Wilcer, Neal and Zhang, Helin and Zhou, Chao and Cancelo, Gustavo},
	month = mar,
	year = {2024},
	pages = {013305},
}

@article{stefanazzi_qick_2022,
	title = {The {QICK} ({Quantum} {Instrumentation} {Control} {Kit}): {Readout} and control for qubits and detectors},
	volume = {93},
	issn = {0034-6748, 1089-7623},
	shorttitle = {The {QICK} ({Quantum} {Instrumentation} {Control} {Kit})},
	url = {https://pubs.aip.org/rsi/article/93/4/044709/2849124/The-QICK-Quantum-Instrumentation-Control-Kit},
	doi = {10.1063/5.0076249},
	language = {en},
	number = {4},
	urldate = {2025-09-23},
	journal = {Review of Scientific Instruments},
	author = {Stefanazzi, Leandro and Treptow, Kenneth and Wilcer, Neal and Stoughton, Chris and Bradford, Collin and Uemura, Sho and Zorzetti, Silvia and Montella, Salvatore and Cancelo, Gustavo and Sussman, Sara and Houck, Andrew and Saxena, Shefali and Arnaldi, Horacio and Agrawal, Ankur and Zhang, Helin and Ding, Chunyang and Schuster, David I.},
	month = apr,
	year = {2022},
	pages = {044709},
}

@article{darmawan_practical_2021,
	title = {Practical {Quantum} {Error} {Correction} with the {XZZX} {Code} and {Kerr}-{Cat} {Qubits}},
	volume = {2},
	issn = {2691-3399},
	url = {https://link.aps.org/doi/10.1103/PRXQuantum.2.030345},
	doi = {10.1103/PRXQuantum.2.030345},
	number = {3},
	urldate = {2025-05-15},
	journal = {PRX Quantum},
	author = {Darmawan, Andrew S. and Brown, Benjamin J. and Grimsmo, Arne L. and Tuckett, David K. and Puri, Shruti},
	month = sep,
	year = {2021},
	pages = {030345},
}

@article{frattini_3-wave_2017,
	title = {3-wave mixing {Josephson} dipole element},
	volume = {110},
	issn = {0003-6951, 1077-3118},
	url = {https://pubs.aip.org/apl/article/110/22/222603/33972/3-wave-mixing-Josephson-dipole-element},
	doi = {10.1063/1.4984142},
	number = {22},
	urldate = {2025-09-15},
	journal = {Applied Physics Letters},
	author = {Frattini, N. E. and Vool, U. and Shankar, S. and Narla, A. and Sliwa, K. M. and Devoret, M. H.},
	month = may,
	year = {2017},
	pages = {222603},
}

@article{frattini_observation_2024,
	title = {Observation of {Pairwise} {Level} {Degeneracies} and the {Quantum} {Regime} of the {Arrhenius} {Law} in a {Double}-{Well} {Parametric} {Oscillator}},
	volume = {14},
	issn = {2160-3308},
	url = {https://link.aps.org/doi/10.1103/PhysRevX.14.031040},
	doi = {10.1103/PhysRevX.14.031040},
	number = {3},
	urldate = {2025-04-22},
	journal = {Physical Review X},
	author = {Frattini, Nicholas E. and Cortiñas, Rodrigo G. and Venkatraman, Jayameenakshi and Xiao, Xu and Su, Qile and Lei, Chan U. and Chapman, Benjamin J. and Joshi, Vidul R. and Girvin, S. M. and Schoelkopf, Robert J. and Puri, Shruti and Devoret, Michel H.},
	month = sep,
	year = {2024},
	pages = {031040},
}

@article{grimm_stabilization_2020,
	title = {Stabilization and operation of a {Kerr}-cat qubit},
	volume = {584},
	issn = {0028-0836, 1476-4687},
	url = {https://www.nature.com/articles/s41586-020-2587-z},
	doi = {10.1038/s41586-020-2587-z},
	number = {7820},
	urldate = {2025-07-02},
	journal = {Nature},
	author = {Grimm, A. and Frattini, N. E. and Puri, S. and Mundhada, S. O. and Touzard, S. and Mirrahimi, M. and Girvin, S. M. and Shankar, S. and Devoret, M. H.},
	month = aug,
	year = {2020},
	pages = {205--209},
}

@article{krantz_quantum_2019,
	title = {A quantum engineer's guide to superconducting qubits},
	volume = {6},
	issn = {1931-9401},
	url = {https://pubs.aip.org/apr/article/6/2/021318/570326/A-quantum-engineer-s-guide-to-superconducting},
	doi = {10.1063/1.5089550},
	number = {2},
	urldate = {2025-05-15},
	journal = {Applied Physics Reviews},
	author = {Krantz, P. and Kjaergaard, M. and Yan, F. and Orlando, T. P. and Gustavsson, S. and Oliver, W. D.},
	month = jun,
	year = {2019},
	pages = {021318},
}

@article{puri_stabilized_2019,
	title = {Stabilized {Cat} in a {Driven} {Nonlinear} {Cavity}: {A} {Fault}-{Tolerant} {Error} {Syndrome} {Detector}},
	volume = {9},
	issn = {2160-3308},
	shorttitle = {Stabilized {Cat} in a {Driven} {Nonlinear} {Cavity}},
	url = {https://link.aps.org/doi/10.1103/PhysRevX.9.041009},
	doi = {10.1103/PhysRevX.9.041009},
	number = {4},
	urldate = {2025-04-17},
	journal = {Physical Review X},
	author = {Puri, Shruti and Grimm, Alexander and Campagne-Ibarcq, Philippe and Eickbusch, Alec and Noh, Kyungjoo and Roberts, Gabrielle and Jiang, Liang and Mirrahimi, Mazyar and Devoret, Michel H. and Girvin, S. M.},
	month = oct,
	year = {2019},
	pages = {041009},
}

@article{putterman_hardware-efficient_2025,
	title = {Hardware-efficient quantum error correction via concatenated bosonic qubits},
	volume = {638},
	issn = {0028-0836, 1476-4687},
	url = {https://www.nature.com/articles/s41586-025-08642-7},
	doi = {10.1038/s41586-025-08642-7},
	number = {8052},
	urldate = {2025-05-14},
	journal = {Nature},
	author = {Putterman, Harald and Noh, Kyungjoo and Hann, Connor T. and MacCabe, Gregory S. and Aghaeimeibodi, Shahriar and Patel, Rishi N. and Lee, Menyoung and Jones, William M. and Moradinejad, Hesam and Rodriguez, Roberto and Mahuli, Neha and Rose, Jefferson and Owens, John Clai and Levine, Harry and Rosenfeld, Emma and Reinhold, Philip and Moncelsi, Lorenzo and Alcid, Joshua Ari and Alidoust, Nasser and et. al.},
	month = feb,
	year = {2025},
	pages = {927--934},
}

@article{xu_tailored_2023,
	title = {Tailored {XZZX} codes for biased noise},
	volume = {5},
	issn = {2643-1564},
	url = {https://link.aps.org/doi/10.1103/PhysRevResearch.5.013035},
	doi = {10.1103/PhysRevResearch.5.013035},
	number = {1},
	urldate = {2025-05-15},
	journal = {Physical Review Research},
	author = {Xu, Qian and Mannucci, Nam and Seif, Alireza and Kubica, Aleksander and Flammia, Steven T. and Jiang, Liang},
	month = jan,
	year = {2023},
	pages = {013035},
}

@phdthesis{nicholas_e_frattini_three-wave_2021,
author = {Frattini,Nicholas E.},
year = {2021},
title = {Three-Wave Mixing in Superconducting Circuits: Stabilizing Cats with SNAILs},
journal = {ProQuest Dissertations and Theses},
pages={177},
note={Copyright - Database copyright ProQuest LLC; ProQuest does not claim copyright in the individual underlying works; Last updated - 2023-03-08},
isbn={9798790626906},
language={English},
url={https://ezproxy.lib.utexas.edu/login?url=https://www.proquest.com/dissertations-theses/three-wave-mixing-superconducting-circuits/docview/2631644043/se-2},
school = {Yale University},
}

@article{venkatraman_driven_2024,
	title = {A driven {Kerr} oscillator with two-fold degeneracies for qubit protection},
	volume = {121},
	copyright = {https://creativecommons.org/licenses/by/4.0/},
	issn = {0027-8424, 1091-6490},
	url = {https://pnas.org/doi/10.1073/pnas.2311241121},
	doi = {10.1073/pnas.2311241121},
	number = {24},
	urldate = {2025-07-18},
	journal = {Proceedings of the National Academy of Sciences},
	author = {Venkatraman, Jayameenakshi and Cortiñas, Rodrigo G. and Frattini, Nicholas E. and Xiao, Xu and Devoret, Michel H.},
	month = jun,
	year = {2024},
	note = {Publisher: Proceedings of the National Academy of Sciences},
}

@article{gertler_protecting_2021,
	title = {Protecting a bosonic qubit with autonomous quantum error correction},
	volume = {590},
	issn = {0028-0836, 1476-4687},
	url = {https://www.nature.com/articles/s41586-021-03257-0},
	doi = {10.1038/s41586-021-03257-0},
	number = {7845},
	urldate = {2025-04-17},
	journal = {Nature},
	author = {Gertler, Jeffrey M. and Baker, Brian and Li, Juliang and Shirol, Shruti and Koch, Jens and Wang, Chen},
	month = feb,
	year = {2021},
	pages = {243--248},
}

@article{hajr_high-coherence_2024,
	title = {High-{Coherence} {Kerr}-{Cat} {Qubit} in {2D} {Architecture}},
	volume = {14},
	issn = {2160-3308},
	url = {https://link.aps.org/doi/10.1103/PhysRevX.14.041049},
	doi = {10.1103/PhysRevX.14.041049},
	number = {4},
	urldate = {2025-05-14},
	journal = {Physical Review X},
	author = {Hajr, Ahmed and Qing, Bingcheng and Wang, Ke and Koolstra, Gerwin and Pedramrazi, Zahra and Kang, Ziqi and Chen, Larry and Nguyen, Long B. and Jünger, Christian and Goss, Noah and Huang, Irwin and Bhandari, Bibek and Frattini, Nicholas E. and Puri, Shruti and Dressel, Justin and Jordan, Andrew N. and Santiago, David I. and Siddiqi, Irfan},
	month = nov,
	year = {2024},
	pages = {041049},
}

@article{frattini_squeezed_2024,
	title = {The squeezed {Kerr} oscillator: spectral kissing and phase-flip robustness},
	volume = {14},
	issn = {2160-3308},
	shorttitle = {The squeezed {Kerr} oscillator},
	url = {http://arxiv.org/abs/2209.03934},
	doi = {10.1103/PhysRevX.14.031040},
	number = {3},
	urldate = {2025-07-18},
	journal = {Physical Review X},
	author = {Frattini, Nicholas E. and Cortiñas, Rodrigo G. and Venkatraman, Jayameenakshi and Xiao, Xu and Su, Qile and Lei, Chan U. and Chapman, Benjamin J. and Joshi, Vidul R. and Girvin, S. M. and Schoelkopf, Robert J. and Puri, Shruti and Devoret, Michel H.},
	month = sep,
	year = {2024},
	note = {arXiv:2209.03934 [quant-ph]},
}

@article{reglade_quantum_2024,
	title = {Quantum control of a cat-qubit with bit-flip times exceeding ten seconds},
	volume = {629},
	issn = {0028-0836, 1476-4687},
	url = {http://arxiv.org/abs/2307.06617},
	doi = {10.1038/s41586-024-07294-3},
	number = {8013},
	urldate = {2025-05-14},
	journal = {Nature},
	author = {Réglade, Ulysse and Bocquet, Adrien and Gautier, Ronan and Cohen, Joachim and Marquet, Antoine and Albertinale, Emanuele and Pankratova, Natalia and Hallén, Mattis and Rautschke, Felix and Sellem, Lev-Arcady and Rouchon, Pierre and Sarlette, Alain and Mirrahimi, Mazyar and Campagne-Ibarcq, Philippe and Lescanne, Raphaël and Jezouin, Sébastien and Leghtas, Zaki},
	month = may,
	year = {2024},
	note = {arXiv:2307.06617 [quant-ph]},
	pages = {778--783},
}

@article{terhal_towards_2020,
	title = {Towards scalable bosonic quantum error correction},
	volume = {5},
	issn = {2058-9565},
	url = {https://iopscience.iop.org/article/10.1088/2058-9565/ab98a5},
	doi = {10.1088/2058-9565/ab98a5},
	number = {4},
	urldate = {2025-05-14},
	journal = {Quantum Science and Technology},
	author = {Terhal, B M and Conrad, J and Vuillot, C},
	month = oct,
	year = {2020},
	pages = {043001},
}

@article{ding_quantum_2024,
  title={Quantum control of an oscillator with a Kerr-cat qubit},
  author={Ding, Andy Z and Brock, Benjamin L and Eickbusch, Alec and Koottandavida, Akshay and Frattini, Nicholas E and Corti{\~n}as, Rodrigo G and Joshi, Vidul R and de Graaf, Stijn J and Chapman, Benjamin J and Ganjam, Suhas and others},
  journal={Nature Communications},
  volume={16},
  number={1},
  pages={5279},
  year={2025},
  publisher={Nature Publishing Group UK London}
}

@article{ofek_demonstrating_2016,
  title={Demonstrating Quantum Error Correction that Extends the Lifetime of Quantum Information},
  author={Ofek, Nissim and Petrenko, Andrei and Heeres, Reinier W and Reinhold, Philip and Leghtas, Zaki and Vlastakis, Brian and Liu, Yehan and Frunzio, Luigi and Girvin, Steven M and Jiang, Liang and others},
  journal={Nature},
  volume={536},
  pages={441--445},
  year={2016}
}

@article{mirrahimi_dynamically_2014,
	title = {Dynamically protected cat-qubits: a new paradigm for universal quantum computation},
	volume = {16},
	copyright = {http://iopscience.iop.org/info/page/text-and-data-mining},
	issn = {1367-2630},
	shorttitle = {Dynamically protected cat-qubits},
	url = {https://iopscience.iop.org/article/10.1088/1367-2630/16/4/045014},
	doi = {10.1088/1367-2630/16/4/045014},
	number = {4},
	urldate = {2025-04-17},
	journal = {New Journal of Physics},
	author = {Mirrahimi, Mazyar and Leghtas, Zaki and Albert, Victor V and Touzard, Steven and Schoelkopf, Robert J and Jiang, Liang and Devoret, Michel H},
	month = apr,
	year = {2014},
	pages = {045014},
}

\end{document}